\documentclass[aoas]{imsart}

\makeatletter
\def\journal@name{The Annals of Applied Statistics}
\makeatother

\RequirePackage{amsthm,amsmath,amsfonts,amssymb}
\RequirePackage[authoryear]{natbib}
\RequirePackage[colorlinks,citecolor=blue,urlcolor=blue]{hyperref}
\RequirePackage{graphicx}

\usepackage{bbm}
\usepackage{verbatim,url,dsfont}
\usepackage{algorithm,algorithmic}
\usepackage{mathtools,enumitem}
\usepackage{multirow}
\usepackage{array}
\usepackage{booktabs}
\usepackage{nicefrac}

\makeatletter
\newcommand*\rel@kern[1]{\kern#1\dimexpr\macc@kerna}
\newcommand*\widebar[1]{%
  \begingroup
  \def\mathaccent##1##2{%
    \rel@kern{0.8}%
    \overline{\rel@kern{-0.8}\macc@nucleus\rel@kern{0.2}}%
    \rel@kern{-0.2}%
  }%
  \macc@depth\@ne
  \let\math@bgroup\@empty \let\math@egroup\macc@set@skewchar
  \mathsurround\z@ \frozen@everymath{\mathgroup\macc@group\relax}%
  \macc@set@skewchar\relax
  \let\mathaccentV\macc@nested@a
  \macc@nested@a\relax111{#1}%
  \endgroup
}
\makeatother

\DeclareMathOperator*{\minimize}{minimize}
\DeclareMathOperator*{\maximize}{maximize}

\DeclareMathOperator{\st}{subject\,\,to}

\DeclareMathOperator{\Cor}{Cor}

\DeclareMathOperator{\Var}{Var}

\DeclareMathOperator{\diag}{diag}

\def\E{\mathbb{E}}
\def\P{\mathbb{P}}
\def\R{\mathbb{R}}

\def\hbeta{\hat{\beta}}

\def\th{^{\textnormal{th}}}

\def\FigureDir{Figures/Color}

\startlocaldefs
\theoremstyle{plain}

\newtheorem{proposition}{Proposition}[section]

\theoremstyle{definition}

\newcommand{\given}{\, \vert \,}
\newcommand{\US}{U.S.}

\newcommand{\ahcomment}[1]{}
\newcommand{\abcomment}[1]{}
\newcommand{\jgcomment}[1]{}
\newcommand{\dmcomment}[1]{}

\endlocaldefs

\begin{document}

\begin{frontmatter}
\title{Estimating Time-Varying Epidemic Severity Rates with Adaptive
  Deconvolution}
\runtitle{Estimating Time-Varying Epidemic Severity Rates}

\begin{aug}
\author[A]{\fnms{Jeremy} \snm{Goldwasser}\ead[label=e1]{jeremy\_goldwasser@berkeley.edu}}
\author[B]{\fnms{Addison J.} \snm{Hu}\ead[label=e2]{addisonh@cmu.edu}}
\author[C]{\fnms{Alyssa} \snm{Bilinski}\ead[label=e3]{alyssa\_bilinski@brown.edu}}
\author[D]{\fnms{Daniel J.} \snm{McDonald}\ead[label=e4]{daniel.mcdonald@ubc.ca}}
\author[A]{\fnms{Ryan J.} \snm{Tibshirani}\ead[label=e5]{ryantibs@berkeley.edu}}

\runauthor{Goldwasser, Hu, Bilinski, McDonald and Tibshirani}

\address[A]{Department of Statistics, University of California,
  Berkeley\printead[presep={,\ }]{e1,e5}}
\address[B]{Department of Statistics and Machine Learning Department,
  Carnegie Mellon University\printead[presep={,\ }]{e2}}
\address[C]{Departments of Health Policy and Biostatistics, Brown
  University\printead[presep={,\ }]{e3}}
\address[D]{Department of Statistics, University of British
  Columbia\printead[presep={,\ }]{e4}}
\end{aug}

\begin{abstract}
Several key metrics in public health convey the probability that a primary event
will lead to a more serious secondary event in the future. These ``severity
rates'' can change over the course of an epidemic in response to shifting
conditions like new therapeutics, variants, or public health interventions. In
practice, time-varying parameters such as the case-fatality rate are typically
estimated from aggregate count data. Prior work has demonstrated that
commonly-used ratio-based estimators can be highly biased, motivating the
development of new methods. In this paper, we develop an adaptive deconvolution
approach based on approximating a Poisson-binomial model for secondary events,
and we regularize the maximum likelihood solution in this model with a trend
filtering penalty to produce smooth but locally adaptive estimates of severity
rates over time. This enables us to compute severity rates both retrospectively
and in real time. Experiments based on COVID-19 death and hospitalization data
show that our deconvolution estimator is generally more accurate than the
standard ratio-based methods, and displays reasonable robustness to model
misspecification.
\end{abstract}

\begin{keyword}
\kwd{Severity rates}
\kwd{Case-fatality rate}
\kwd{Deconvolution}
\kwd{Trend filtering}
\kwd{COVID-19}
\kwd{Epidemiology}
\end{keyword}
\end{frontmatter}

\section{Introduction}

Many important public health metrics report the probability that a secondary,
often more severe outcome, follows a given primary event. The most common of
these ``severity rates'' is the case-fatality rate, or CFR \citep{cfr_cao,
nishiuraEx1}. This is sometimes treated as a proxy for the infection-fatality
rate, or IFR, which is generally more challenging to compute \citep{timevar_ifr,
lancet_ifr}. Other examples of interest are based on different types of primary
events, for example hospitalizations \citep{HFR_linelist3} and wastewater
shedding \citep{wastewater_cases}.

It is common to treat severity rates as stationary over time \citep{ghani,
  jewell2007nonparametric, reich2012estimating, lancet_controversial}. This
assumes that the population-level probability of developing a secondary
event, given a primary event, is constant throughout the entire epidemic. In
reality, however, severity rates often vary in time---due to evolving factors
like the introduction of vaccines \citep{cfr_vaccines}, improved treatments  
\citep{paxlovid}, and the emergence of new variants 
\citep{cfrs_delta_omicron}. For example, \citet{cfrs_by_variant} estimates the  
original COVID-19 variant to have a 3.6\% CFR globally, compared to 2.0\% for
the Delta variant, and 0.7\% for Omicron.  

In an ideal setting, severity rates could be calculated directly using detailed
line-list or claims data that track outcomes of individual patients
\citep{HFR_linelist3, cfr_line_list, HFR_linelist1, HFR_linelist2}. However,
during fast-moving epidemics like COVID-19, comprehensive real-time tracking on
a large geographic scale has generally been impractical. As a result, severity
rates are commonly estimated using aggregate counts of primary and secondary 
events over time. Indeed, CFR estimates based on aggregate case and death counts
became ubiquitous during the COVID-19 pandemic. These appeared not only in
scientific studies \citep{LIU2023100350, germany, cfr_lancet}, but also in major
news outlets such as The Atlantic \citep{atlantic}, Wall Street Journal
\citep{wsj}, and New York Times \citep{nyt}.  

In this paper, we focus on estimating the severity rate, defined as
\begin{equation}
\label{eq:severity}
p_t = \P(\text{secondary event will occur} \given \text{primary event at $t$}), 
\end{equation}
at each given point in time $t$. Severity rates can be estimated
\emph{retrospectively}, where data collected after time $t$ is used to estimate  
$p_t$. Retrospective analyses can be valuable, as they can provide insights into 
epidemic dynamics amidst changing conditions. Often of greater interest,
however, is estimating severity rates in \emph{real time}, which means only data
available up until $t$ may be used to estimate $p_t$. This is far more
challenging, and expanding \eqref{eq:severity} helps explain why: we have 
\smash{$p_t = \sum_{k=0}^{\infty} \P(\text{secondary event occurs at $t+k$}
  \given \text{primary event at $t$})$}, which depends on epidemic events (of 
potentially changing likelihood) in the future, yet real-time estimators for
$p_t$ may only use data through $t$, and none afterwards.

Common methods for constructing real-time severity rate estimates are based on
taking ratios of primary and secondary event counts. These are described later
in Section \ref{sec:methods}. In past work \citep{goldwasser}, we demonstrated
that such ratio-based estimators can be subject to large (and predictable)
statistical biases. Common retrospective estimators take on similar forms, and
generally exhibit similar biases.

In this work, we propose an approach that overcomes the limitations of these
methods. Our method has less bias than the existing ratio-based estimators, as
it is based on performing deconvolution in a model that explicitly encodes
time-varying severity rates in order to generate secondary events from primary 
events. At its core, our deconvolution model stems from a Poisson-binomial 
characterization of the relationship between primary and secondary event counts.  
Since this model is underspecified, we use regularization---specifically, a form
of trend filtering---to smooth out the estimated severity rates, while
maintaining the ability to adapt to potentially abrupt changes in the underlying
signal (thanks to the local adaptivity of trend filtering). In the real-time
case, we use additional regularization at the right tail of the sequence,
originally developed in \citet{Jahja2022}. Our work expands on
\citet{fusedlasso}, which proposes to estimate severity rates using a similar
but more restricted framework, based on squared loss and total variation
penalties.  

We validate our methodology on a broad range of experiments. These experiments
start with real data on COVID-19 hospitalizations during the pandemic, and
hypothetical hospitalization-fatality rate (HFR) curves that we craft based on
associating the major four variants (original strain, Alpha, Delta, and Omicron)
with fixed but data-driven HFRs. We then simulate COVID-19 deaths using a
Poisson-binomial (or beta-binomial, in cases where overdispersion is present)
model which uses observed hospitalization counts and hypothetical HFR curves.
Tuning all hyperparameters with cross-validation, our deconvolution method
yields around 15\% lower MAE than the convolutional ratio estimator in the  
retrospective case. This improvement increases to 55\% when comparing to the
lagged ratio, arguably the standard choice in current practice
\citep{yuan2020monitoring, timevar_ifr, horita2022global, lagged_chinese,
  LIU2023100350}. The benefits in the real-time case are similarly strong. Our
method continues to outperform ratio-based methods under various degrees of
model misspecification. Finally, the same qualitative benefits are observed on 
real COVID-19 deaths, as we describe shortly in the subsection below.

The rest of this article is structured as follows. After previewing and
motivating our methodology with a real data application comparing HFR estimates
in Section \ref{sec:real-data}, we develop in Section \ref{sec:stat-model} the
Poisson-binomial model, and corresponding approximations used for
deconvolution. In Section \ref{sec:estimators}, we introduce our approaches for
estimating severity rates, first in the retrospective setting and then in
real-time. In Section \ref{sec:setup}, we describe our experimental setup, and
in Section \ref{sec:results}, we analyze its results. Section
\ref{sec:discussion} concludes with a discussion.

\subsection{Motivating application: time-varying HFR in COVID-19}
\label{sec:real-data}  

In this subsection, we compare our deconvolution approach to the existing
ratio-based methods in estimating the COVID-19 hospitalization-fatality rate (HFR) 
over the pandemic in the state of Pennsylvania. The HFR is an example of
a severity rate \eqref{eq:severity} of key interest in public health, where primary
events are hopsitalizations and secondary events are deaths. Instead of
estimating a single number for the entire COVID-19 pandemic, we estimate the HFR
as a function of time---allowing the HFR to change in response to changing
underlying conditions, such as the introduction of new therapeutics, or the 
emergence of new variants.      

We consider two different settings in which we estimate the HFR curve: the
retrospective setting, where we use data that was only available in hindsight to
estimate the HFR at each time $t$ as best as possible; and the real-time
setting, where we limit ourselves to data available at each time $t$ in order to
estimate the HFR at $t$. Our proposed method, which performs regularized
deconvolution in a likelihood model, is described in Sections \ref{sec:retro}
and \ref{sec:real-time} for in the retrospective and real-time settings,
respectively; existing ratio-based methods are described in Section
\ref{sec:methods}.

The data used in our experiments in this subsection are from standard public
health reporting pipelines. For primary events, we use daily COVID-19 hospital
admissions, as coordinated by the National Healthcare Safety Network (NHSN). For
secondary events, we use different datasets for the retrospective and real-time
cases. In retrospect, we use daily COVID-19 death counts,\footnote{This data is 
  only available at the weekly level, hence we imputed daily deaths by sampling
  from a multinomial distribution with the weekly NCHS death total as the number
  of trials and uniform probabilities (all equal to $1/7$).}  
as collected by the National Center for Health Statistics (NCHS). In real-time,
we use daily COVID-19 deaths, as assimilated by Johns Hopkins University
(JHU). NCHS deaths were only available weeks after each event in question, while
JHU published provisional death counts in real-time. These data sources are
described in more detail in Section \ref{sec:data-generation}. 

The estimators compared here all depend on a delay distribution, representing
the (stochastic) transition from hospitalizations to deaths. We take this to be
a discretized gamma distribution (or a point mass, for the lagged ratio method),
whose mean is chosen separately in the retrospective and real-time cases by  
maximizing cross-correlation between the relevant observables, in the same
manner as described later in Section \ref{sec:data-generation}. 
All methods have hyperparameters, which we tune via cross-validation and the
one standard error (1se) rule, as described in Section \ref{sec:design}. 
We favor the 1se rule because the min rule qualitatively tends to undersmooth, and even with this smoothness-favoring tuning, the ratio-based methods still exhibit the instability highlighted below.
For the
deconvolution method, we choose quadratic order ($m=2$) regularization in the
retrospective case, because this yields a smooth HFR curve, and constant order 
($m=0$) in the real-time case, because this displays the most stable performance 
across our experimental suite to come later.  

\begin{figure}[t]
\centering
\includegraphics[width=0.95\textwidth]{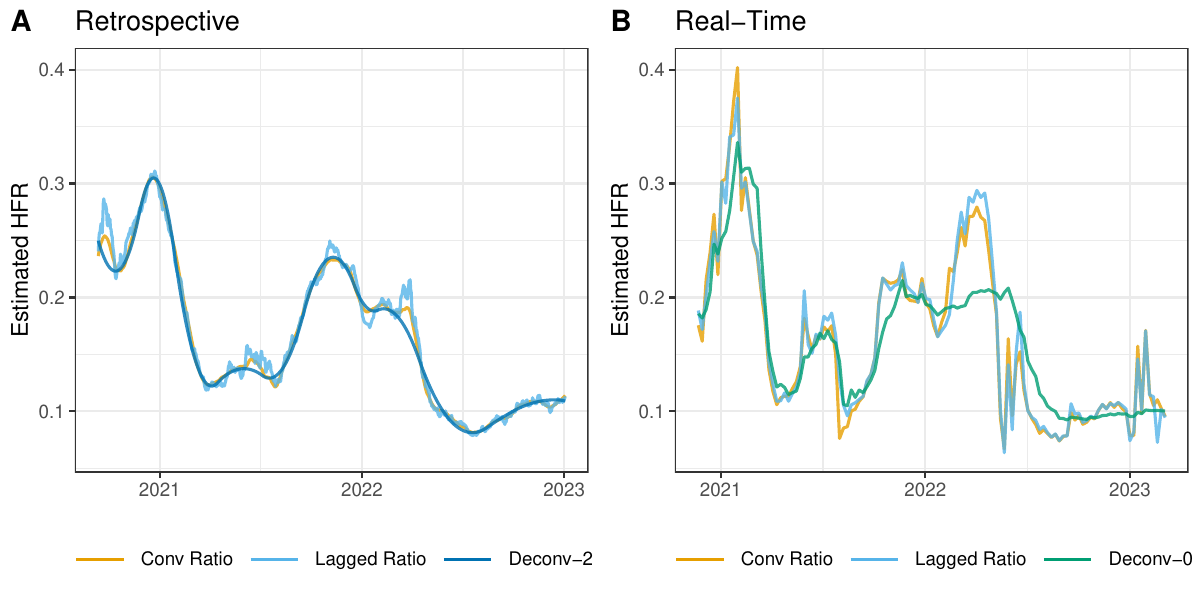}
\caption{HFR estimates on real COVID-19 data in Pennsylvania.}
\label{fig:real-data}
\end{figure}

Figure \ref{fig:real-data} displays the retrospective and real-time severity
rate estimates. The high-level summary---across both settings (retrospective and 
real-time) and all methods---is as follows. We see HFRs peaking at around 30\%
during the surge around the beginning of 2021, then falling below 15\% as the
surge ended. Conditions remained relatively steady with the Alpha variant in
spring 2021, then climbed rapidly with the Delta variant later that year. They
dipped to around 20\% as the Omicron variant swept through the state in winter
2022. As the wave passed, severity rates fell sharply, bottoming out around
10\%. 

Comparing estimates to each another, the deconvolution estimates demonstrate
notable stability relative to the ratio-based estimates. This is true in both
retrospective and real-time cases, but especially pronounced in the real-time
case, displayed in panel B. In this case, the ratio estimates spike to about
40\% at the peak of the winter wave in early 2021, while the deconvolution
estimate stays near 30\% (consistent with all estimates in the retrospective
case). Moreover, the ratio estimates suspiciously \emph{increase} as the Omicron
wave passes in April 2022, whereas the deconvolution estimate holds steady over
the same period, before smoothly declining (again broadly consistent with the
retrospective estimates).

There is no ground truth HFR in these real data experiments against which we can
measure our estimates, to ultimately determine which is more accurate. However,
the suspicious behavior of the ratio-based methods in Figure \ref{fig:real-data}
is consistent with the prior analysis in \citet{goldwasser}.
For example, that work reveals how the real-time estimators are upwardly biased during periods of sharp decline in the primary counts, including early 2022. 
Furthermore, in the Sections \ref{sec:setup} and \ref{sec:results} of 
this paper, we carry out semi-synthetic experiments which start from real
primary events (COVID-19 hospitalizations), and then simulate secondary events 
(COVID-19 deaths) from custom models which are tailored to match both the trend
and noise levels in observed secondary events (NCHS and JHU deaths in the
retrospective and real-time settings, respectively). 
This allows us to recreate the qualitative differences between the estimators seen in Figure \ref{fig:real-data} and quantify the degree to which the proposed deconvolution approach improves over ratio-based alternatives.

\section{Statistical model}\label{sec:stat-model}

In this section, we present a model which relates secondary to primary event
counts, via severity rates, and describe an approximate likelihood suitable for
inference.   

\subsection{Exact likelihood}\label{sec:exact-lik}

At a time point $t$, let $X_t$ and $Y_t$ be the aggregate numbers of
primary and secondary events, respectively. In the context of CFR, for example,
$X_t$ would be the number of new cases at time $t$, and $Y_t$ the number of new 
deaths. Throughout we consider discrete integer-valued time points such as 
$t=1,2,3,\dots$, but allow negative time values, for notational simplicity (for  
example, when indexing the set of time points that precede a given time $t$, as 
we do below).   

In general, the number of secondary events at time $Y_t$ can be expressed as a
sum of indicator functions, where each indicator represents whether a given
primary in the past event resulted in a secondary event at $t$. Using cases and
deaths as primary and secondary events, for concreteness, observe that 
\begin{equation}
\label{eq:secondary_incidence}
Y_t = \sum_{k=0}^{\infty} \sum_{i=1}^{X_{t-k}} \mathds{1}\{\text{case $i$
  occurring at time $t-k$ dies at $t$}\}. 
\end{equation}
The secondary incidence time series can be understood probabilistically by
noting that the indicator functions are Bernoulli random variables. Applying the
definition of conditional probability,   
\begin{align}\label{eq:expectation}
&\E[\mathds{1}\{\text{case $i$ at $t-k$ dies at $t$}\}] \\ 
&\quad = \P(\text{dies at $t$} \given \text{case at $t-k$})\nonumber \\
&\quad = \P(\text{death occurs $\cap$ dies at $t$} 
  \given \text{case at $t-k$})\nonumber \\ 
&\quad = \P(\text{death occurs} \given \text{case at $t-k$}) \cdot
 \P(\text{dies at $t$} \given \text{case at $t-k$, death occurs}).\nonumber
\end{align}
The first term of the final line is the severity rate $p_{t-k}$, from
\eqref{eq:severity}. Meanwhile, we define the second term to be the $k\th$
element of the \emph{delay distribution} at time $t-k$. Formally, define  
\begin{equation}
\pi_k^{(t)} = \P(\text{dies at $t+k$} \given \text{case at $t$, death occurs}) 
\end{equation}
as the $k\th$ element of the delay distribution at time $t$, where
\smash{$\sum_{k=0}^\infty\pi_k^{(t)} = 1$}. 
Thus by definition the expectation \eqref{eq:expectation} is equal to $p_{t-k} \pi_k^{(t-k)}$, so we may write   
\begin{equation}
\label{eq:indicator_bernoulli}
\mathds{1}\{\text{case $i$ occurring at time $t-k$ dies at $t$}\} \sim
\mathrm{Bernoulli}(p_{t-k} \pi_k^{(t-k)}).
\end{equation}
It is reasonable to assume that these Bernoulli random variables are
independent. Each represents a different patient, whose outcomes should not affect
one another. 
An exception could arise if severity rates are calculated within a very small, resource-constrained group---for example, if one patient's death frees up hospital resources for others, marginally improving their prognosis \citep{nursing}.
In general, however, severity rates are
typically assessed over a broad population, at the resolution of counties or 
states or even countries---not single hospitals or nursing homes. Henceforth, we make the
assumption that individual patient outcomes are independent. 
This assumption of independent Bernoulli outcomes is shared by all other works on severity rates we compare to \citep{UKpaper,germany,ghani,timevar_ifr}.

Under this assumption, note that \smash{$Y_t \given X_{s\leq t}$} follows a
\emph{Poisson-binomial} distribution, where we abbreviate \smash{$X_{s\leq t} =
  \{X_s : s \leq t\}$}. The Poisson-binomial generalizes the binomial
distribution, in that it represents a sum of independent Bernoulli random variables that do not necessarily share the same success probability. 
In the context of severity rates, each Bernoulli trial is whether an individual patient has a secondary event time $t$, as expressed in \eqref{eq:secondary_incidence}. 
Note by \eqref{eq:indicator_bernoulli} that all individuals with primary events on the same dates have the same ``success'' probability; this is $\pi_{k}^{(t-k)}p_{t-k}$ for all $X_{t-k}$ trials at time $t-k$.
Thus we may concisely write  
\begin{equation}
\label{eq:pb}
Y_t \given X_{s\leq t} \sim \mathrm{PoissonBinomial} \Big( 
\underbrace{\pi_0^{(t)}p_{t},\dots,\pi_0^{(t)}p_{t}}_{X_{t}}, \, 
\underbrace{\pi_1^{(t-1)}p_{t-1},\dots,\pi_1^{(t-1)}p_{t-1}}_{X_{t-1}}, \,
\dots \Big),
\end{equation}
where the underbraces indicate the number of times each success probability is
repeated. A useful fact to record will be the mean and variance of this
distribution:
\begin{align}
\label{eq:pb_mean}
\mu_t &= \E[Y_t \given X_{s\leq t}] = \sum_{k=0}^\infty X_{t-k} \pi_k^{(t-k)}    
p_{t-k}, \\
\label{eq:pb_var}
\sigma_t^2 &= \Var(Y_t\given X_{s\leq t}) = \sum_{k=0}^\infty
  X_{t-k}\pi_k^{(t-k)} p_{t-k} (1-\pi_k^{(t-k)} p_{t-k}).  
\end{align}
which follow by linearity of expectation and independence.

\subsection{Approximate likelihood}\label{sec:approx}

\paragraph*{Approximations to the Poisson-binomial}

The Poisson-binomial likelihood is generally intractable for large counts. 
As with the binomial distribution, for the probability mass function at 
$Y_t = y_t$, it is necessary to consider all \smash{$\sum_{k=0}^\infty X_{t-k}$}
choose $y_t$ combinations of patients that may yield this realized number of
secondary events. For the Poisson-binomial, however, these combinations do not
share the same probability of occurring, since the Bernoulli variates in
\eqref{eq:pb} have different success probabilities. As a result, evaluating the
probability mass function requires enumerating all products of $y_t$ Bernoulli 
probabilities. This combinatorial explosion is computationally prohibitive
unless the total number of primary events is very small, which is not the case
in our application, where we are concerned with COVID-19 cases or
hospitalizations for \US\ states, or across the whole nation.

Fortunately, different approximations exist for the Poisson-binomial
distribution. A natural choice is the Poisson distribution with rate $\mu_t$
from \eqref{eq:pb_mean}. This has nonnegative, albeit unbounded, support. It has
variance \smash{$\sum_{k=0}^\infty X_{t-k} \pi_k^{(t-k)} p_{t-k}$}, which is
larger than $\sigma_t^2$ in \eqref{eq:pb_var}, but this difference matters
little in practice, at least in our applications of interest. For example, using
numbers from the national experiments we run in Section \ref{sec:setup}, the
Poisson variance is only bigger by about 1.28\%. More formally,
\citet{barbour1984poisson} showed that the Poisson approximation to the
Poisson-binomial incurs small error when the Bernoulli random variables have low
success rates. This is generally true for severity rates in epidemics, where the
probabilities \smash{$\pi_k^{(t-k)} p_{t-k}$} are small. By the results of
\citet{barbour1984poisson}, the total variation distance between the
Poisson-binomial in \eqref{eq:pb} and its corresponding Poisson approximation
(with rate $\mu_t$) is upper bounded by
\begin{equation}
\sum_{k=0}^{\infty}(\pi_k^{(t-k)}p_{t-k})^2.
\end{equation}
Again employing data from our \US\ simulation, the above bound is only 
$1.69 \times 10^{-5}$. The negligible TV distance justifies approximating the distribution of $Y_t \mid X_{s \leq t}$ as Poisson.

Alternatively, \citet{pb_theory} studies a normal approximation to the
Poisson-binomial, with mean $\mu_t$ and variance $\sigma_t^2$. Their empirical
and theoretical results show the normal approximation more faithfully captures
the upper tail of the cumulative density function than the Poisson. However, it 
struggles with the lower tail when the mean is low, since it includes negative
support.  

\paragraph*{Weak dependence at successive time points}

Conditional on past counts of primary events, we have shown that secondary event
counts follow a Poisson-binomial distribution. Strictly speaking, these
secondary event counts need not be conditionally independent at successive time
points, since they are defined over a common group of individuals. For example,
if case $i$ from $t-k$ dies at time $t$, then this same case cannot die at time
$t+1$. Fortunately, the next result demonstrates that the counts from successive
time steps are weakly dependent, under a simplifying assumption of equal
variance.  

\begin{proposition}\label{prop:corr}
Assume \smash{$\Var(Y_t \given X_{s\leq t}) = \Var(Y_{t+1}\given X_{s\leq
  t+1})$}. Under the Poisson-binomial model described in
\eqref{eq:secondary_incidence}, \eqref{eq:indicator_bernoulli}, \eqref{eq:pb},
it holds that
\begin{equation}
\Cor(Y_t, Y_{t+1} \given X_{s\leq t+1}) \in \bigg[\frac{-\max_{k \geq 0}
  \pi_k^{(t-k)} p_{t-k}}{1-\max_{k \geq 0} \pi_k^{(t-k)} p_{t-k}}, \, 0 \bigg].    
\end{equation}
\end{proposition}

Appendix A of the Supplementary Material \citep{goldwasser2026supp} contains the proof of this result. Given reasonably
long-tailed delay distributions with low severity rates, each
\smash{$\pi_k^{(t-k)} p_{t-k}$} will be quite small. Plugging in numbers
corresponding to the national data considered in Section \ref{sec:setup}, the 
lower bound on the correlations from Proposition \ref{prop:corr} is roughly
-0.018. Computing the actual correlations from our simulated data, the lowest
observed value is -0.012. As these numbers are so low, it seems reasonable to
approximate the distribution of \smash{$Y_t \given X_{s\leq t}$} as being
independent over different time points $t$, for the sake of estimation.   

While our likelihood characterization for severity rates is novel, several works in compartmental models and reproduction numbers also assume aggregate counts on different timesteps are independent \citep{cori2013new, rtestim, cauchemez2008likelihood, wallinga_teunis}. 


\paragraph*{Joint distribution and MLE}

We have established that \smash{$Y_t \given
X_{s \leq t}$} may be approximated by Poisson or Gaussian distributions, which are nearly conditionally independent 
as $t$ varies. Next, we characterize the joint distribution over $t$, leaving regularization to the next
section. Let $p$ denote the vector of severity rates, which has coordinate $p_t$
at time $t$.  By approximating the law of \smash{$Y_t \given X_{s\leq t}$} as
Poisson with rate $\mu_t = \mu_t(p)$ as in \eqref{eq:pb_mean}, and assuming 
independence of these conditional distributions over $t$, we arrive at the
following maximum likelihood problem: 

\begin{align}
&\maximize_{0 \preceq p \preceq 1} \; \prod_t \frac{\mu_t(p)^{Y_t}e^{-\mu_t(p)}}  
  {Y_t!} \\
\label{eq:mle-pois}
\iff &\minimize_{0 \preceq p \preceq 1} \; \sum_t \bigg[ \bigg(\sum_{k=0}^\infty     
  X_{t-k}\pi_k^{(t-k)}p_{t-k}\bigg) - Y_t\log\bigg(\sum_{k=0}^\infty
  X_{t-k}\pi_k^{(t-k)}p_{t-k}\bigg)\bigg],  
\end{align}
where the notation $0 \preceq z \preceq 1$ denotes elementwise constraints.  
This optimization problem falls in the class of \emph{Poisson linear
  inverse} problems, well-studied in statistical physics and image processing 
\citep{richardson1972bayesian, lucy1974iterative, dupe2011linear,
  rond2016poisson}. For a Poisson linear inverse problem to be computationally
tractable, the linear operator being applied to the optimization variable must
have nonnegative entries. That is indeed the case here, as each of the terms 
\smash{$X_{t-k} \pi_k^{(t-k)}$} are nonnegative.      

We note that the identity-link parameterization used in \eqref{eq:mle-pois}
contrasts with Poisson regression, which is a generalized linear model that may 
be more familiar to many statisticians. In this model, the log of the rate is a
linear function of the features, and we would learn parameters $\beta$ such that   
\smash{$\log\mu_t = \sum_{k=0}^\infty X_{t-k}\pi_k^{(t-k)} \beta_{t-k}$}. However,  
having estimated $\beta$, there would be no unique way to recover the severity
rates $p$ from 
\begin{equation}
\log \bigg( \sum_{k=0}^\infty X_{t-k}\pi_k^{(t-k)}p_{t-k} \bigg) =
\sum_{k=0}^\infty X_{t-k}\pi_k^{(t-k)}\beta_{t-k}.
\end{equation}

In addition to the Poisson model described above, we also studied and 
implemented a severity rate estimator using the normal approximation. While both    
estimators performed roughly similarly well in our experiments, the Poisson
model was simpler (note that the severity rates $p$ appear in the variance in 
\eqref{eq:pb_var}, which complicates the Gaussian approximation), and had
marginally higher accuracy. For that reason, we only provide a high-level
account of the Gaussian methods in the main text, delegating further discussion
to Appendix B of the Supplementary Material.

\paragraph*{Estimating the delay distribution}
In most applications, the oracle delay distribution is unknown. In lieu of
\smash{$\pi^{(t)}$}, a plug-in estimate \smash{$\hat\pi^{(t)}$} can be used
instead, which typically has a finite support, placing all of its mass on
$[0,d]$, where $d < \infty$. In other words, this assumes that no secondary
events occur after $d$ time steps (and turns all infinite sums in
\eqref{eq:mle-pois} from $k=0$ to $\infty$ into finite sums from $k=0$ to 
$d$). In any case, whether or not the support of the delay distribution is
finite, there will always be more severity rates than secondary event times, and
so the maximum likehood problem \eqref{eq:mle-pois} is underspecified. The  
estimators we propose in Section \ref{sec:estimators} will thus use
regularization.  

Estimating the delay distributions \smash{$\pi^{(t)}$} is an entire line of work
in and of itself. Many approaches use line lists which contain the times of
primary and secondary events, other approaches may be based on parametric
approximations
whose parameters are based on small observational studies or even chosen based
on epidemiological literature. 
Popular choices of parametrizations include discretized gamma, Weibull, and log-normal distributions.
For a thorough account of existing methods, we
refer the reader to \citet{delay_distrs}.

\section{Severity rate estimators}\label{sec:estimators}

We now present our proposed severity rate estimators for both the retrospective
and real-time settings. We focus on the Poisson likelihood, with Appendix
B of the Supplementary Material covering the Gaussian case. Some form of regularization is
necessary to compute likelihood-based severity rates, due to the fact that the
maximum likelihood problem (either Poisson or Gaussian) is underspecified. While
many forms of regularization would suffice to make the problem well-specified,
we use a form which aligns with the smoothness considerations underlying
severity rates, discussed in detail below. We then present our retrospective and
real-time estimators. 

For our methods to work well, the approximate likelihood outlined in Section \ref{sec:approx} must match the true data-generating process reasonably well. 
To review, this model assumes that individuals have independent outcomes, the Poisson (or Gaussian) approximates the Poisson-binomial accurately, $Y_{t}$ and $Y_{t+1}$ are conditionally independent, and the delay distribution is estimated well. 
In addition, the following estimators assume the actual severity rates are smooth. The nature of the smoothness assumption depends on the order of trend filtering regularization, introduced next.

\subsection{Trend filtering regularization}\label{sec:tf}

Trend filtering is a nonparametric estimation technique, which models the mean
function of univariate data as a piecewise polynomial where the order (e.g.,
linear or cubic) is user-specified \citep{og_tf}. While this is reminiscent of
the smoothing spline, trend filtering can adjust more responsively to the local 
level of smoothness in the underlying signal \citep{Tibshirani2014} by
adaptively selecting the locations---also called knots---at which the fitted  
polynomial changes. Local adaptivity is important for estimating severity rates,
as these may remain constant for long stretches of time when epidemic conditions
do not change substantially, before changing rapidly in response to new
conditions.     

For an integer order $m\geq 0$, trend filtering in its original form (for
nonparametric regression) models the mean vector $\theta \in \R^n$ over $n$ data
points as a piecewise polynomial, of degree $m$. It does so by minimizing a loss
term while penalizing the differences of $\theta$ of order $m+1$. These
differences can be compactly represented by multiplying $\theta$ by a difference
operator $D^{(m+1)}$. This can be viewed as the discrete analog of a derivative 
operator, and is defined recursively as
\begin{equation}
D^{(m+1)} = D^{(1)}\,D^{(m)}\in\mathbb{R}^{(n-m-1)\times n}
\end{equation}
where $D^{(1)}$ is the first difference operator
\begin{equation}
D^{(1)} =
\begin{bmatrix}
-1 & 1 & 0 & \cdots & 0 & 0\\
0 & -1 & 1 & \cdots & 0 & 0\\
\vdots & & \ddots & \ddots & & \vdots\\
0 & 0 & 0 & \cdots & -1 & 1
\end{bmatrix}
\,\in\,\R^{(n-m-1)\times (n-m)}.
\end{equation}
Trend filtering applies a penalty based on the $\ell_1$ norm, $\lambda
\|D^{(m+1)} \theta\|_1$, for a tuning parameter $\lambda \geq 0$. To give
examples, for the first three orders $m=0,1,2$, the trend filtering penalties
are
\[
\begin{aligned}
\|D^{(1)} \theta\|_1 &= \sum_{i=1}^{n-1} |\theta_{i+1} - \theta_i|, \\
\|D^{(2)} \theta\|_1 &= \sum_{i=1}^{n-2} |\theta_{i+2} - 2\theta_{i+1} +
  \theta_i|, \\
\|D^{(3)} \theta\|_1 &= \sum_{i=1}^{n-3} |\theta_{i+3} - 3\theta_{i+2} +
  3\theta_{i+1} - \theta_i|.
\end{aligned}
\]
Since the $\ell_1$ norm induces sparsity, trend filtering solutions
\smash{$\hat\theta$} have the property that many elements of the differenced
vector \smash{$D^{(m+1)}\hat\theta$} will be exactly zero \citep{genlasso}, with
generally more zeros for larger $\lambda$ values. The nonzero elements
correspond to knots, which are adaptively chosen based on the data. Between the
knots, the solution traces out a degree $m$ polynomial; different segments of 
\smash{$\hat\theta$} may possess more or less smoothness, depending on how close
the knots are to each other. We refer to \citet{Tibshirani2014} for more
details, and to \citet{tibshirani2022divided} for more discussion of the broader
context and relation to splines.

The special case of $m=0$ produces a piecewise constant fit and is known as the
fused lasso \citep{og_fused_lasso} or total variation denoising
\citep{rudin1992nonlinear}, and has a longer history of study. Recently, 
\citet{fusedlasso} proposed to use a fused lasso penalty to estimate severity
rates in a regression framework with squared loss. The methods detailed below
can be understood as a generalization of their work, encompassing higher-order
trend filtering penalties, a Poisson loss which stems from approximating the
Poisson-binomial likelihood underlying secondary event generation (Section
\ref{sec:stat-model}), and added tail regularization to tame real-time 
estimation (Section \ref{sec:real-time}, below). That said, the squared loss
used in \citet{fusedlasso} is close to the Gaussian approximation we describe in 
Appendix B of the Supplementary Material.

\subsection{Retrospective deconvolution}\label{sec:retro}

\paragraph*{Estimator}

Our retrospective method appends a trend filtering penalty to the Poisson
likelihood approximation derived in Section \ref{sec:approx}. In particular,
starting from \eqref{eq:mle-pois}, we use a plug-in estimate
\smash{$\hat\pi^{(t)}$} for the delay distribution, with finite support $[0,d]$,
and append a trend filtering penalty of order $m \geq 0$, which yields 
\begin{equation}
\label{eq:tf-pois}
\minimize_{0 \preceq p \preceq 1} \; \sum_t \bigg[\bigg( \sum_{k=0}^d   
X_{t-k}\hat\pi_k^{(t-k)}p_{t-k} \bigg) - Y_t\log\bigg( \sum_{k=0}^d 
X_{t-k}\hat\pi_k^{(t-k)}p_{t-k} \bigg)\bigg] + \lambda\|D^{(m+1)}p\|_1. 
\end{equation}
We denote the vector of restrospective severity rate
estimates, obtained by solving \eqref{eq:tf-pois}, by
\smash{$\hat{p}^{\text{rs}}$}. 
Problem \eqref{eq:tf-pois} is convex, and can be readily optimized with standard software. Our implementation uses CVXR \citep{cvxr}, the R interface to the disciplined convex programming framework CVX \citep{gb08}. CVXR supports a variety of back-end solvers, of which we use the Clarabel optimizer for its efficiency on cone programs \citep{clarabel}. We discuss the computational cost of this solver in Appendix E of the Supplementary Material. Alternatively, one could adapt the specialized alternating direction method of multipliers optimizers developed for trend filtering \citep{ramdas2016fast,Jahja2022}, but we do not pursue this in the current paper.

The sum in \eqref{eq:tf-pois} is over all secondary event times. Let $N_Y$
denote the number of such event times. Note that there are $N_Y+d$ estimated
severity rates, starting at $d$ time steps before the first secondary event
time. Not all of these estimates are equally reliable. In this model, the
severity rate $p_t$ contributes to secondary events in the $d$ time steps which
follow. Away from the left or right boundary of second event times, assuming
$N_Y$ is sufficiently large relative to $d$, these secondary events are all
observed. As a result, the estimate \smash{$\hat{p}^{\text{rs}}_t$} is likely to
be more stable, as it contributes to (and is hence informed by) many elements of
the loss. 

On the other hand, severity rate estimates will be less stable when $t$ is near
the left or right boundaries of observations. In the most extreme edge cases,
the first and last severity rates only affect a single secondary count. This
data scarcity issue persists throughout the $d$ time steps of each tail. As they
contribute little to the loss, severity rate estimates at the tails are in
general subject to greater variability. To account for this, we recommend (and
implement) a burn-in and burn-out period for retrospective estimation. That is,
while $N_Y+d$ severity rates are obtained, it is prudent to ignore the values
near the boundaries.

\paragraph*{Cross-validation}

The hyperparameter $\lambda \geq 0$ in \eqref{eq:tf-pois} controls the degree of
smoothness exhibited by \smash{$\hat{p}^{\text{rs}}$}. In an extreme case 
($\lambda \to \infty$), the solution will take the form of a global polynomial
of degree $m$. At the other extreme ($\lambda \to 0$), it will be a highly
volatile piecewise polynomial, having a knot at each possible time point. We
rely on $K$-fold cross-validation for selecting $\lambda$, defining folds in a
structured way that respects the time dependence in our retrospective estimation
problem. In each of the $K$ folds, secondary incidence $Y_t$ at every 
$K\th$ time step $t$ is held out from the loss, reducing the number of training 
samples by a factor of $1/K$. For fold $j \in \{1,\dots,K\}$, denoting by
\smash{$\hat{p}^{\text{rs},j}(\lambda)$} the result of optimizing the
corresponding version of \eqref{eq:tf-pois} with samples withheld, we record 
the mean absolute error (MAE) from reconvolution over the validation set $V_j$,  
\begin{equation}
\text{MAE}_j(\lambda) = \frac{1}{|V_j|} \sum_{t \in V_j} \bigg| Y_{t} -
\sum_{k=0}^d X_{t-k}\hat\pi_k^{(t-k)} \hat{p}^{\text{rs},j}_{t-k}(\lambda)
\bigg|.  
\end{equation}
The cross-validation error for the given $\lambda$ is the average across the $K$
folds:
\begin{equation}
\text{MAE}(\lambda) = \frac{1}{K} \sum_{j=1}^K \text{MAE}_j(\lambda).
\end{equation}
This process is iterated over a grid of $\lambda$ values, the largest of which
should ideally produce an estimate with no knots. We denote this value by
\smash{$\lambda_\text{max}$}, and derive its form in Appendix
C of the Supplementary Material. Observe that any \smash{$\lambda \geq \lambda_\text{max}$}
will result in an estimate which is a global polynomial. From the
cross-validation error curve, we select the value $\lambda^*$ of the tuning 
parameter with the lowest error, and then use this value to estimate severity
rates using the whole time series. This is a standard strategy, sometimes called
the ``min rule.'' An alternative that is useful when the cross-validation error
curve is flat around its minimum is the ``1se rule'' \citep{hastie2009elements},
which selects the largest $\lambda$ whose cross-validation error is within one
standard error of the minimum value.  

Since fewer training samples are used when tuning $\lambda$ in cross-validation,
we make a slight adjustment to \eqref{eq:tf-pois} in order to ensure that it
balances the loss and regularization terms on a common scale. Specifically,
whenever solving \eqref{eq:tf-pois} (either in cross-validation iterations or in
full, on the whole time series), we normalize each sum by its number of
summands. 

The order of trend filtering, $m \geq 0$, may also be tuned using the data. We
can compare cross-validation errors across all $\lambda$ and $m$, and select the
pair $(\lambda^*,m^*)$ with the lowest error. As we we typically restrict our
attention to $m \in \{0,1,2\}$---resulting in piecewise constant, linear, and 
quadatic trends, respectively---such additional tuning over $m$ does not present 
much of an additional computational burden. Alternatively, the user may want 
to handpick $m \in \{0,1,2\}$ based on qualitative considerations, to reflect
the desired shape of the severity rate curve: piecewise constant, allowing for
jump discontinuities; piecewise linear, allowing for sharp turnaround points; or
piecewise quadratic, allowing for smoother evolution.

\subsection{Real-time deconvolution}\label{sec:real-time}

\paragraph*{Estimator}

Denote by $T$ the time through which data is available and at which one seeks 
to estimate the current severity rate $p_T$. To be clear, in this real-time
estimation problem, no data which becomes available after $T$ may be used, since
it does not yet exist. The most basic adaption of the retrospective approach in
the previous subsection to the current real-time case would be to solve
\eqref{eq:tf-pois}, where the sum is restricted to $t \leq T$. However, here the
severity rate $p_T$ only contributes to a single data point $Y_T$ in the loss,
and this leads to a much greater degree of variability in estimating $p_T$
compared to the retrospective case. We will thus use extra regularization to
temper such tail variability. This itself introduces a potential tradeoff,
gaining stability but reducing adaptivity.

The regularization techniques we employ are inspired by \citet{Jahja2022}, who 
studied deconvolution in the context of nowcasting infections that will
eventually appear as case reports. The additional regularization comes in two 
parts. The first part is to impose a constraint on the tail values of the
severity rate curve, which prevents overfitting to the most
recent secondary counts. In particular, we constrain the differences of order
$m+1$ of the last $m$ severity rates to be zero, enforcing these rates  
to adhere to a polynomial trend of degree $m$ (preventing a knot from occuring
near the tail). This constraint differs only lightly from that in
\citet{Jahja2022}, who used a natural spline constraint. This constrains the 
tails to be a polynomial of degree $(m-1)/2$; note that our proposal applies to
all values of $m$, whereas natural splines are only defined for odd $m$.  

The second part adds a regularization term to the criterion which is a kind
of tapered smoothing penalty, penalizing the squared differences $p_{t-1}-p_t$
in adjacent severity rates with successively decreasing weight as $t$ moves away
from $T$. Such weights are chosen based on how much mass is captured by the
delay distribution between primary and secondary events. Specifically, we assign
$(p_{t-1}-p_t)^2$ a weight \smash{$w_t = 1/\hat{F}^{(T)}(T-t)$} for $t \geq
T-d$, and $w_t = 0$ for $t < T-d$, where \smash{$\hat{F}^{(T)}(k) =
  \sum_{j=0}^k \hat\pi^{(T)}_j$} is the CDF of \smash{$\hat\pi^{(T)}$} at 
$k$. Thus, the longer the tail of the delay distribution, the more we seek to
smooth out severity rates in the recent past.    

Putting these two sources of regularization together leads to the following  
optimization, which defines our real-time severity rate estimates at $T$:   
\begin{equation}
\begin{alignedat}{2}
\label{eq:tf-pois-rt}
&\minimize_{0 \preceq p \preceq 1} \;\; && \sum_{t \leq T} \bigg[\bigg( \sum_{k=0}^d
X_{t-k}\hat\pi_k^{(t-k)}p_{t-k} \bigg) - Y_t\log\bigg( \sum_{k=0}^d
X_{t-k}\hat\pi_k^{(t-k)}p_{t-k} \bigg)\bigg] \\
& && \qquad + \lambda\|D^{(m+1)}p\|_1 + \gamma \|W D^{(1)}p\|_2^2 \\
&\st \;\; && \sum_{k=0}^{m+1} (-1)^k \binom{m+1}{k} p_{T - m - 1 + k} = 0,
\end{alignedat}
\end{equation}
where \smash{$W = \diag(0,\dots,0,\sqrt{w_{T-d}},\dots,\sqrt{w_T})$} is a
diagonal matrix whose only nonzero elements are the final $d+1$ diagonal 
entries. Denoting by \smash{$\hat{p}^{\text{rt}}$} the vector of real-time
severity rate estimates obtained by solving \eqref{eq:tf-pois-rt}, the real-time
estimate of $p_T$ is its last element \smash{$\hat{p}^{\text{rt}}_T$}.

In practice, it is common for epidemic data streams to be revised after initial
data is released \citep{reinhart2021open}. This means that, from one time $T$ to
the next, the whole sequences of primary and secondary counts used in
real-time estimation may be updated (rather than new primary and secondary
counts simply being appended). To reflect this, it helps to introduce some
additional notation: for time points $s \geq t$, denote by \smash{$X^{(s)}_t$}
the version of primary incidence at time $t$ which was available as of time $s$,
and similarly \smash{$Y^{(s)}_t$} for secondary incidence. Therefore, in
practice, we would substitute each pair $X_t,Y_t$ of primary and secondary
counts in \eqref{eq:tf-pois-rt} with \smash{$X^{(T)}_t,Y^{(T)}_t$}, the versions
of these counts available at $T$. 

\paragraph*{Cross-validation}

We tune the hyperparameters $\lambda, \gamma \geq 0$ in \eqref{eq:tf-pois-rt}
with a two-stage approach, again inspired by \citet{Jahja2022}. First, setting
$\gamma=0$, tune $\lambda$ with $K$-fold validation, exactly as in the
retrospective case described previously. Then, fixing $\lambda = \lambda^*$ at
the chosen value from cross-validation, we tune $\gamma$ with a
rolling-validation (also called forward-validation) procedure. This works as
follows: for each $s = T-M,\dots,T-1$, we solve \eqref{eq:tf-pois-rt} using 
only data up through time $s$, and use the resulting severity rate estimates
\smash{$\hat{p}^{\text{rt},s}(\gamma)$} to form a prediction
\smash{$\hat{Y}_{s+1}(\gamma)$} of $Y_{s+1}$ by reconvolution. The 
rolling-validation error for the given  
$\gamma$ is then 
\begin{equation}
\text{MAE}(\gamma) = \frac{1}{M} \sum_{s=T-M}^{T-1} \big| \hat{Y}_{s+1}(\gamma) 
- Y_{s+1} \big|. 
\end{equation}
This process is iterated over a grid of $\gamma$ values. As with
cross-validation, we can select a value $\gamma^*$ according to the ``min rule''  
(the $\gamma$ with minimum MAE), or alternatively, the ``1se rule'' (the largest 
$\gamma$ whose MAE lies within one standard error of the minimum). Finally, the 
real-time problem is solved at hyperparameter values $\lambda^*$ and $\gamma^*$
on all available data, from which we extract \smash{$\hat{p}^{\text{rt}}_T$}.
This is summarized in Algorithm \ref{alg:fv}.  
Figure 3 in Appendix E of the Supplementary Material plots deconvolution estimates at varying levels of $\gamma$, highlighting its ability to significantly alter tail predictions. 

It is worth making two further remarks. First, as in the retrospective case,
before running either cross-validation or forward-validation, the likelihood and
regularizer terms in \eqref{eq:tf-pois-rt} are placed on a common scale by
dividing each by their number of summands. Second, the order of trend filtering
$m$ can again be tuned using the two-stage approach, based on minimizing the 
forward-validation error. It can instead be chosen based on qualitative
considerations, as discussed in the retrospective subsection.  

\renewcommand{\algorithmicrequire}{\textbf{Input:}}
\renewcommand{\algorithmicensure}{\textbf{Output:}}

\begin{algorithm}[htb]
\caption{Tuning $\lambda,\gamma$ by two-stage validation}
\begin{algorithmic}[1]\label{alg:fv}
\REQUIRE Candidate sets of values $\Lambda,\Gamma$ for $\lambda,\gamma$; primary
and secondary incidence $X_t,Y_t$, for $t \leq T$; number of cross-validation
folds $K$ and number of forward-validation steps $M$ 
\ENSURE Selected $\lambda^*, \gamma^*$ 

\FOR{each $\lambda \in \Lambda$}
\STATE Use $K$-fold cross-validation for problem \eqref{eq:tf-pois-rt} with
$\gamma = 0$, to record $\text{MAE}(\lambda)$ 
\ENDFOR
\STATE Select $\lambda^*$ by min or 1se rule
\FOR{each $\gamma \in \Gamma$}
\FOR{$s = T - M$ to $T - 1$}
\STATE Solve \eqref{eq:tf-pois-rt} with given $\gamma$, $\lambda = \lambda^*$,  
and $T=s$ to yield $\vphantom{\sum_{k=0}^d} \hat{p}^{\text{rt},s}(\gamma)$
\STATE Linearly extrapolate $\vphantom{\sum_{k=0}^d}
\hat{p}^{\text{rt},s}_{s+1}(\gamma)$ from
$\hat{p}^{\text{rt},s}_s(\gamma)$ and 
$\hat{p}^{\text{rt},s}_{s-1}(\gamma)$    
\STATE Compute prediction $\hat{Y}_{s+1}(\gamma) = \sum_{k=0}^d X_{s+1-k} 
  \hat\pi^{(s)}_k \hat{p}^{\text{rt},s}_{s+1-k}(\gamma)$
\ENDFOR
\STATE Record $\text{MAE}(\gamma) = \frac{1}{M} \sum_{s=T-M}^{T-1}
|\hat{Y}_{s+1}(\gamma) - Y_{s+1}|$
\ENDFOR
\STATE Select $\gamma^*$ by min or 1se rule
\STATE \textbf{return} $\lambda^*,\gamma^*$
\end{algorithmic}
\end{algorithm}

\section{Experimental setup}\label{sec:setup}

\subsection{Semi-synthetic data generation}
\label{sec:data-generation}

To evaluate our methods against existing benchmarks, we generated realistic
secondary incidence data using ground truth severity rates, whose computation is 
described below. These simulations are intended to mimic true COVID-19 deaths in
the \US\ over the pandemic. The severity rate we targeted throughout our
analysis was hence the hospitalization-fatality rate (HFR).   

\paragraph*{Real data resources}

Our simulations used real hospitalization counts as the primary incidence
data. In the first several years of the pandemic, COVID-19 hospital admissions
were reported daily in real time to the Department of Health and Human Services 
\citep{HHS2023}. This process was coordinated by the National Healthcare Safety
Network, thus we refer to the aggregate daily hospital admissions as NHSN 
data. 

From observed NHSN hospitalizations, we simulated COVID-19 deaths across the
\US\ and all 50 states, as well as for both the retrospective and real-time
settings. We designed separate simulation models for each of these settings; the
discrepancy here reflects the fact that different counts were available in
real-time versus in retrospect during the pandemic. True weekly death totals
were aggregated in retrospect by the National Center for Health Statistics
(NCHS) \citep{nchs}. NCHS revealed these death totals well after the week in
question. NHSN began comprehensive hospitalization reporting in summer 2020, and
NCHS stopped reporting deaths in early 2023. Our retrospective simulations cover
this 2.5-year window.

In real-time, Johns Hopkins University (JHU) published provisional daily death
counts. These data bear a subtle yet important distinction from those of
NCHS---JHU deaths reflect not the true number of deaths which occurred each day,
but rather the number of new deaths \textit{reported} on a given day. Often,
these deaths were reported days or even weeks after they occurred. As a result,
the delay distribution which underlies the JHU data has higher mean than that
for NCHS. The JHU counts were also considerably noisier, due at least in part to
reporting idiosyncrasies. Our real-time simulations cover the same 2.5-year
window, from summer 2020 to early 2023, consistent with the retrospective case. 

We use the Epidata API \citep{Epidata} to download all of the data described
above.  

\paragraph*{True severity rates}

The severity rates that we used as the ground truth in our simulations were
created using the following variant-based procedure. In each region (state or
\US\ national), we defined a single HFR associated with the four most
significant COVID-19 variants: the original strain, Alpha, Delta, and
Omicron. Using data from CoVariants \citep{Hodcroft2021}, we identified the
dominant period for each variant, based on when it accounted for over half of
cases. Within this window, we computed the HFR as the total number of NCHS
deaths divided by NHSN hospitalizations, offset by two weeks to account for
delays. Finally, we mixed these per-variant HFRs with the variant proportions in
circulation to obtain the final HFR curve: 
\begin{equation}
p_t = \sum_v c_t^v \, p^v, \;\; \text{where} \; \sum_v c_t^v = 1 \;
\text{for all $t$}.
\end{equation}
Here the sums are over variants $v$, with $p^v$ denoting the HFR for variant
$v$, and $c_t^v$ denotes the proportion of variant $v$ in circulation at time
$t$ (calculated again from CoVariants data). 

\paragraph*{Delay distributions}

While our severity rate estimators allow for nonstationary delay distributions
\smash{$\pi^{(t)}$}, for the sake of simplicity our simulations use a constant delay \smash{$\pi^{(t)} =
  \pi$}, for all $t$. 
  In reality, delay distributions are known to vary over time \citep{Ward2021}, though our evaluation of model performance under misspecified delays partially addresses this limitation.
  Moreover, this simplification is consistent with standard practice in estimating severity rates \citep{UKpaper,fusedlasso} and reproduction numbers \citep{wallinga_teunis, cori2013new, Chitwood2022}.
  
  Setting $d=60$ days, we fit
discrete gamma distributions in each region parameterized by
heuristically-chosen means and variances. The means were estimated by maximizing
the cross-correlation between NHSN hospitalizations and deaths, NCHS deaths in
the retrospective case and JHU in real-time. The standard deviations were set to
90\% of the means, motivated by empirical findings on typical
hospitalization-to-death delays \citep{UKdelay}. Denoting by $F_\gamma$ the
cumulative distribution function (CDF) of the corresponding gamma distribution,
we define the delay distribution by \smash{$\pi_k \propto F_\Gamma(k+1) -
  F_\Gamma(k)$}, where these values are normalized to sum to 1.  

\paragraph*{Noise models}

\begin{figure}[tbp]
\centering
\includegraphics[width=0.95\textwidth]{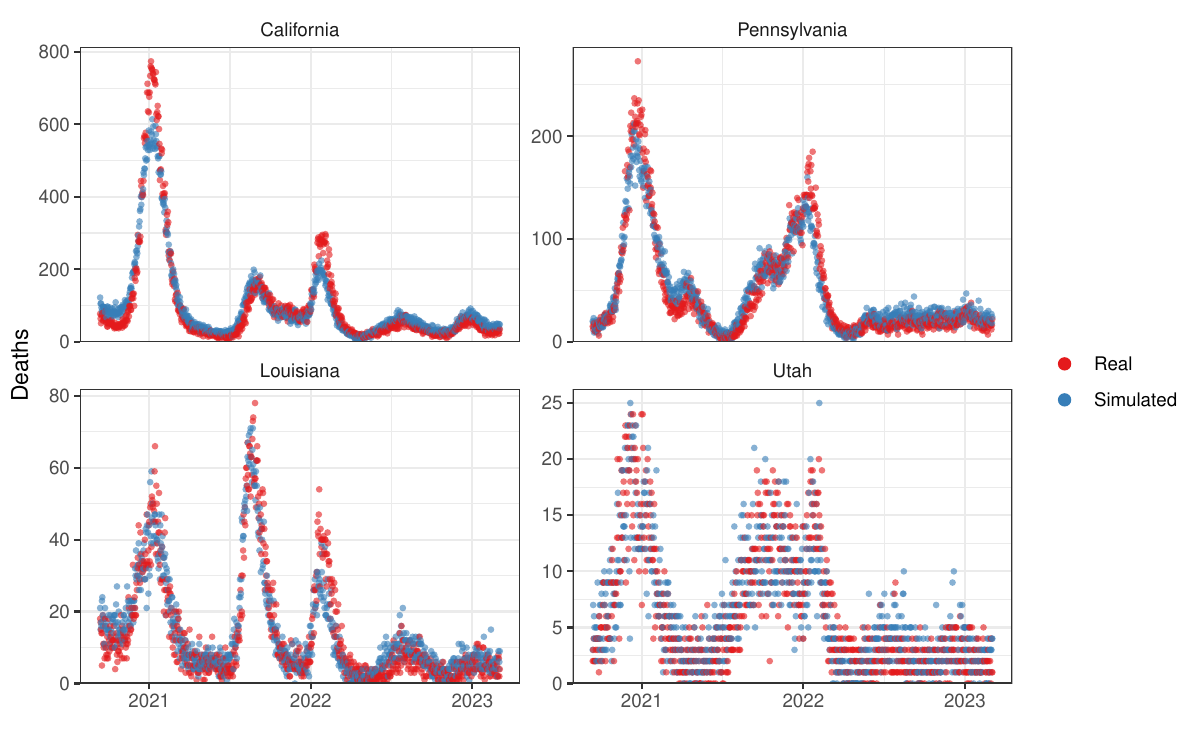}
\caption{Real (NCHS) and simulated (Poisson-binomial) daily deaths. (The
  NCHS data is itself weekly, but here we have subsampled it to the daily level
  to match the time resolution of our simulated data.)}      
\label{fig:sim_vs_real_retro}

\bigskip
\includegraphics[width=0.95\textwidth]{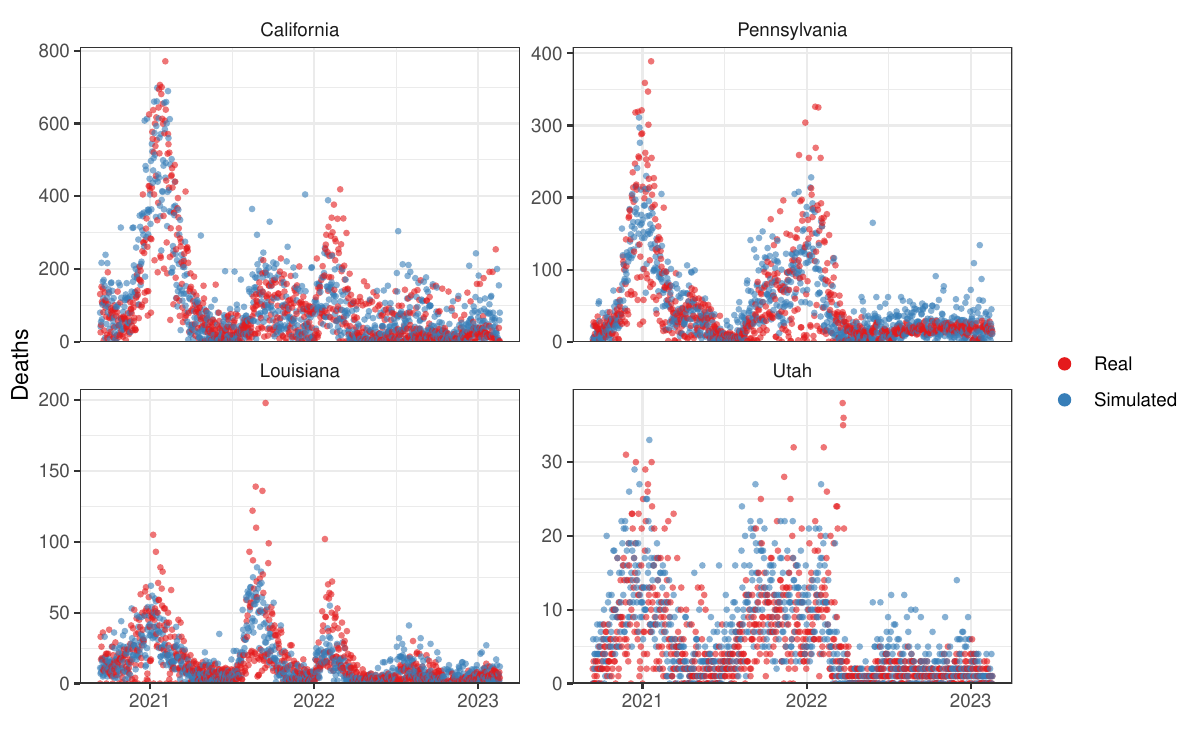} 
\caption{Real (JHU) and simulated (beta-binomial) daily deaths. (The JHU deaths
  displayed here are ``finalized'', meaning that the history has been revised
  after the end of the reporting period. They have been preprocessed as
  described in Appendix D.1 of the Supplementary Material.)}     
\label{fig:sim_vs_real_rt}
\end{figure}

Using the NHSN hospitalizations, variant-based severity rates, and delay 
distributions as described above, we then simulate daily deaths from the
following noise models.  
\begin{itemize}
\item For the retrospective simulation, we generated deaths from a
  Poisson-binomial model; this qualitatively matches the variance of death
  counts from NCHS, as shown in Figure \ref{fig:sim_vs_real_retro}.   
\item For the real-time simulation, we generated deaths from a beta-binomial 
  distribution, which reflects the fact that JHU deaths are more overdispersed
  than what the Poisson-binomial can accommodate. As we describe below, we first
  estimated the amount of overdispersion, then simulated deaths with the given
  variance. The results qualitatively match the dispersion of deaths counts from 
  JHU, see Figure \ref{fig:sim_vs_real_rt}.   
\end{itemize}

To estimate the amount of dispersion in each region (50 states and \US\
overall), we fit a quasi-Poisson regression to cleaned JHU death
counts. Appendix D.1 of the Supplementary Material details our methodology. This regression
computes a coefficient \smash{$\hbeta$} which encodes the amount of
overdispersion present in the JHU death data, in the given region. To
accommodate such overdispersion in our simulation model for deaths, we then use
a beta-binomial model, described next.  

The beta-binomial distribution is a popular tool for modeling count data with
dispersion \citep{betabinom}. For our problem, in a given region with dispersion
coefficient \smash{$\hbeta$}, we can use the standard mean-rho parameterization
to define the beta-binomial at each time $t$, setting      
\begin{equation}
\label{eq:beta-binom}
M_t = \frac{\mu_t}{\sum_{k=0}^d X_{t-k}}, \;\; \text{and} \;\; 
\rho_t = \bigg[ \frac{\hbeta \sigma_t^2}{\mu_t\big(1-\mu_t / \sum_{k=0}^d    
X_{t-k}\big)}-1 \bigg]\bigg[ \frac{1}{\sum_{k=0}^d X_{t-k}-1} \bigg],   
\end{equation}
where $\mu_t, \sigma_t^2$ are as defined in \eqref{eq:pb_mean},
\eqref{eq:pb_var}. Appendix D.2 of the Supplementary Material gives the details behind this
calculation.   

Figures \ref{fig:sim_vs_real_retro} and \ref{fig:sim_vs_real_rt} qualitatively
assess the goodness-of-fit of our simulation models by plotting simulated 
and real data for four regions of varying sizes. While the curves do not overlap 
precisely (we do not have access to the true HFR in simulations), visual
inspection shows that the noise levels are relatively comparable.      

\subsection{Methods considered}\label{sec:methods}

In the retrospective and real-time simulations, we ran Poisson deconvolution
with trend filtering penalties of orders $m = 0,1,2$, which deliver piecewise
constant, linear, and quadratic estimates, respectively. We further considered
tuning the order $m$ of trend filtering regularization itself, by
cross-validation in the retrospective case, and forward-validation in real-time.

We compared these deconvolution methods to the standard ratio-based estimators
of severity rates, which we will often refer to as to ``benchmark'' methods. The
precise formulations for these methods differ slightly between the retrospective
and real-time settings; in all cases, we allow the methods to use smoothed
primary and secondary counts over some window $W \geq 1$, serving as a
hyperparameter (with $W=1$ corresponding to no smoothing). In the real-time
setting, we smooth counts with a trailing window, and in retrospect, we use a
centered window. For the case of primary incidence, define 
\begin{equation} 
\widetilde{X}_t^{\text{rt}} = \frac{1}{W} \sum_{k = 0}^{W-1} X_{t-k},
\;\; \text{and} \;\;
\widetilde{X}_t^{\text{rs}} = \frac{1}{W} \sum_{k = -\lfloor W/2
  \rfloor}^{\lceil W/2 \rceil - 1} X_{t+k},  
\end{equation}
and analogously define \smash{$\widetilde{Y}_t^\text{rt}$} and 
\smash{$\widetilde{Y}_t^\text{rs}$} for secondary incidence. 

With this notation in place we can now define the \emph{lagged ratio}
estimators, arguably the most widely-used estimators of severity rates in
practice. These simply compare primary and secondary events, offset by a lag of
$\ell \geq 0$ time steps. The real-time and retrospective lagged ratios are       
\begin{equation}
\hat{p}_t^\text{rt} =
\frac{\widetilde{Y}_t^\text{rt}}{\widetilde{X}_{t-\ell}^\text{rt}},
 \;\; \text{and} \;\;
\hat{p}_t^\text{rs} =
\frac{\widetilde{Y}_{t+\ell}^\text{rs}}{\widetilde{X}_{t}^\text{rs}},
\end{equation}
respectively. The lag parameter $\ell$ represents the typical duration
between primary and secondary events. 

Two alternative estimators introduced by \citet{UKpaper} (based on earlier work
of \citet{nishiura}) adopt a more faithful approach that uses an estimate of the
delay distribution, \smash{$\hat\pi^{(t)}$}. We call these the
\emph{convolutional ratio} estimators, as the delay distribution convolves 
against trailing primary events to account for secondary incidence. The
real-time convolutional ratio is   
\begin{equation}
\label{eq:conv-rt}
\hat{p}_t^\text{rt} = \frac{\widetilde{Y}_t^\text{rt}}{\sum_{k=0}^d
  \widetilde{X}_{t-\ell}^\text{rt}\hat\pi^{(t-k)}_{k}}.
\end{equation}
The retrospective version tracks the proportion of cases among primary events 
at $t$ who will eventually have secondary events. While secondary event dates  
for individual cases at $t$ are unknown, the relevant secondary count can be 
estimated with a convolutional model. This leads to the retrospective
convolutional ratio:
\begin{align}\label{eq:conv-retro}
\hat{p}_t^\text{rs} = \sum_{k=0}^d
  \frac{\widetilde{Y}_{t+k}^\text{rs}\hat\pi^{(t)}_{k}}{\sum_{j=0}^d
  \widetilde{X}_{t+k-j}^\text{rs}\hat\pi^{(t+k-j)}_{j}}. 
\end{align}
We note that when each distrbution \smash{$\hat\pi^{(t)}$} is a point mass at
$\ell$, the convolutional ratios reduce to the lagged ones. Therefore, the
former may be understood as generalizations of the latter. 

In previous work \citet{goldwasser}, we studied the bias of the real-time lagged
and convolutional ratios, both formally and empirically. Our analysis revealed
that the lagged ratio tends to exhibit large bias, and can both signal
nonexistent surges and fail to detect upticks in severity rate. The
convolutional ratio is generally more robust, but it still can have significant   
bias when the severity rate is changing.      

The retrospective lagged ratio is subject to roughly the same bias as in the 
real-time case: it is effectively the same estimator offset by $\ell$ time 
steps. The retrospective convolutional ratio is somewhat more challenging to
analyze than its real-time counterpart. However, its derivation is based on a
stationarity assumption, and it will still be particularly biased when the
severity rate changes around $t$.  

These ratio estimators all depend on the hyperparameter $W$, which controls the   
length of the smoothing window, and must be tuned. In the retrospective case,
this is tuned via $K$-fold cross-validation, just like $\lambda$ for the
retrospective deconvolution method. In the real-time case, it is tuned via
$M$-step forward-validation, just like $\gamma$ for the real-time deconvolution
method.  

\subsection{Experimental design}\label{sec:design}

In all cases, we examined the use of both the min rule and the 1se rule to tune
hyperparameters using cross- and forward-validation: recall this is $\lambda$
for retrospective deconvolution; $\lambda,\gamma$ for real-time deconvolution;
and $W$ for the benchmark methods, lagged and convolutional ratios, in both
retrospective and real-time settings. In each case, we report results for the
most favorable rule (min or 1se) in terms of MAE, giving each method the full
benefit of the doubt; details will be given in the next section.  We used $K=5$
cross-validation folds and $M=28$ forward-validation steps, throughout.

We evaluated deconvolution and benchmark severity rate estimators on simulated
data across 50 states as well as the \US\ nationally. The estimation window
encompassed a roughly two-year period between 2021 and 2022, omitting a burn-in
and burn-out period. With the exception of the misspecified experiments
described in the final paragraph of this subsection, the deconvolution methods
and the convolutional ratios were given access to the oracle delay distribution, 
thus using \smash{$\hat\pi^{(t)}=\pi$}, for all $t$. The lagged ratios use a lag
$\ell$ equal to the mean of the delay distribution in each region. In all but
the misspecified experiments, the reported results are averaged over 10
replications (10 draws of data from the simulation model).   

To peform deconvolution with trend filtering regularization we solved the
optimization problems \eqref{eq:tf-pois}, \eqref{eq:tf-pois-rt} with Clarabel in
CVXR \citep{clarabel}. The retrospective setting required much less computation
than real-time. For this reason, we also ran deconvolution and the benchmarks
with oracle hyperparameter tuning. This selected the hyperparameters whose
estimated severity rates had the smallest mean absolute error (MAE), in
hindsight. To lessen the computational burden in both the retrospective and
real-time cases, we only computed estimates from each method once every seven
days. In total, this amounted to about 100 estimation dates, over which we
calculated the eventual MAEs to be reported.

Lastly, we recomputed all the estimators under misspecified delay
distributions---still constant over time, but not equal to \smash{$\pi$}. We
considered six different misspecified delay distributions with varying means;
each was still a discrete gamma distribution, whose standard deviation is set to
90\% of its mean. In the retrospective case, we set the means to be 1, 2, and 3
days before and after the true value for each state. For the real-time case,
which had longer delay distributions, the means were offset by 1, 3, and 5 days,
on both sides of the true value. Appendix D.3 of the Supplementary Material provides visualizations
of the misspecified delay distributions for a few states.

\section{Experimental results}\label{sec:results}

We analyze results across the synthetic experiments described in the previous
section.

\subsection{Retrospective analysis}
 
In the retrospective setting, deconvolution largely outperforms the
ratio-based estimators across the \US\ and the 50 states. Table
\ref{tab:retro-cv} summarizes the performance of these methods when all of the
hyperparameters are tuned via cross-validation. As mentioned above in the
experimental design, we consider using both the min rule and 1se rule within 
cross-validation, and for each method we report the results from the rule with
the strongest performance; in the retrospective case, this ends up being the min  
rule for trend filtering, and the 1se rule for the ratio methods. The table
reports, for each method, the MAE averaged over the 51 regions and 10
replicates:      
\begin{equation}
\frac{1}{51} \sum_{r=1}^{51} \bigg (\frac{1}{10}
\sum_{i=1}^{10} \text{MAE}_{ri} \bigg),
\end{equation}
where \smash{$\text{MAE}_{ri}$} is the MAE for region $r$ and replicate $i$. It
also reports the associated standard error:  
\begin{equation}
\sqrt{\frac{1}{51^2}\sum_{r=1}^{51} \frac{1}{10} \hat\sigma^2\big( 
\{\text{MAE}_{ri}\}_{i=1}^{10} \big)},   
\end{equation}
where \smash{$\hat\sigma^2(S)$} is the sample variance of elements in a set
$S$. In the table, ``Deconv-$m$'' denotes deconvolution with trend
filtering regularization of order $m$, and ``Deconv-T'' denotes deconvolution
with the order tuned by cross-validation. The same labeling is used throughout
all  tables and figures.    

As we can see from Table \ref{tab:retro-cv}, all orders of trend filtering yield
more accurate severity rate estimates than the ratio-based methods. The MAE of
the convolutional ratio is roughly $8.3 \times 10^{-3}$ on average,
substantially outperforming the lagged ratio. For deconvolution methods, the
average MAE ranges from $6.7$ to $7.1 \times 10^{-3}$, depending on the order
$m$. This translates to a roughly 12-16\% improvement in MAE over the
convolutional ratio, and roughly 54-56\% over the lagged ratio. Tuning the order
of trend filtering also works well. Moreover, Table 1 in
Appendix E of the Supplementary Material shows that under oracle tuning (with all tuning
parameters chosen to optimize MAE) deconvolution outperforms the benchmarks
by an even wider margin.

\begin{table}[tbp]
\centering
\caption{MAE of methods in retrospective HFR estimation, and the
  associated percentage improvement on the convolutional ratio (CR) and lagged
  ratio (LR). Results are averaged over 51 regions and 10 replications.}    
\label{tab:retro-cv}
\begin{tabular}[t]{@{}lrrrrrr@{}}
\toprule
& Lagged Ratio & Conv Ratio & Deconv-0 & Deconv-1 & Deconv-2 & Deconv-Tuned\\
\midrule
MAE $\times \; 10^3$ & 16.6 ± 0.2 & 8.3 ± 0.1 & 6.7 ± 0.1 & 6.9 ± 0.1 & 7.1 ± 0.1 & 6.9 ± 0.1\\
Improv over CR (\%) & -112.5 ± 2.7 & 0.0 ± 0.0 & 15.2 ± 0.7 & 15.9 ± 0.7 & 12.4 ± 0.8 & 14.1 ± 0.7\\
Improv over LR (\%) & 0.0 ± 0.0 & 47.6 ± 0.6 & 56.2 ± 0.6 & 56.3 ± 0.7 & 54.3 ± 0.7 & 55.6 ± 0.6\\
\bottomrule
\end{tabular}
\end{table}

\begin{figure}[tbp]
\centering
\includegraphics[width=0.82\textwidth]{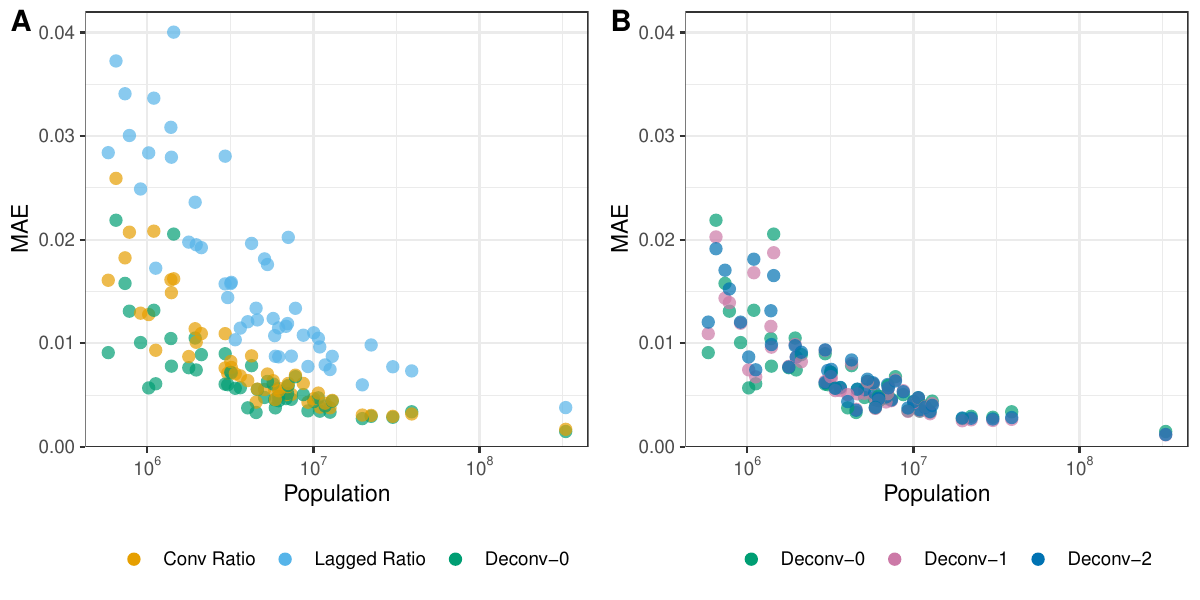}
\caption{MAE versus population size in retrospective HFR estimation, for 51
  regions (\US\ and 50 states). Each point is averaged over 10 replications.}
\label{fig:retrospective-comparison}
\end{figure}

\begin{figure}[tbp]
\centering
\includegraphics[width=0.82\textwidth]{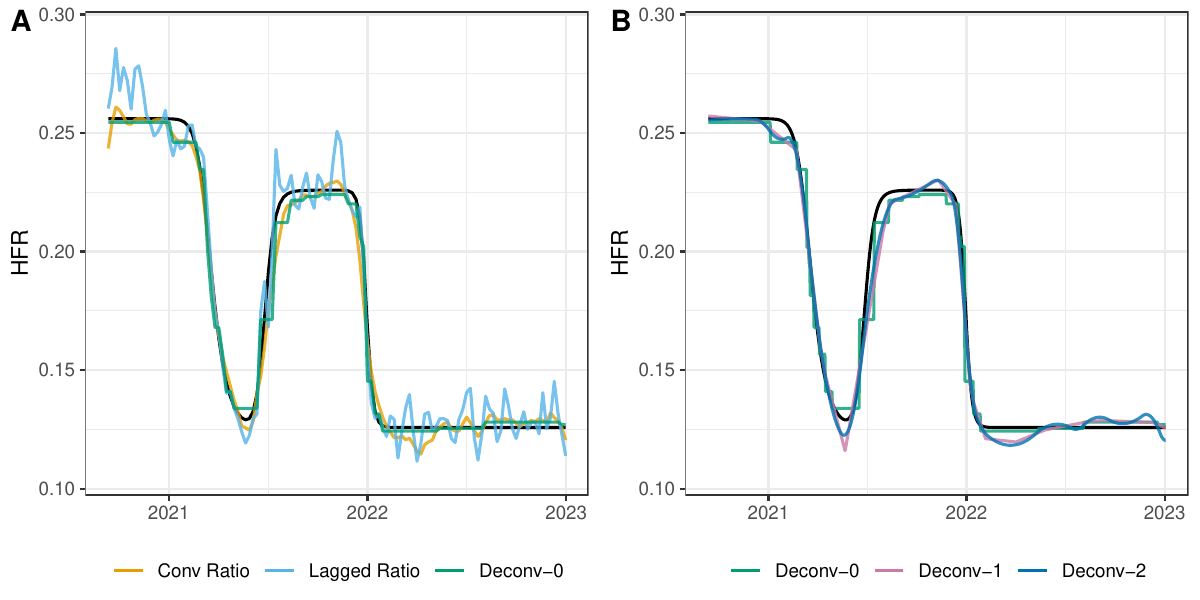}
\caption{Example retrospective HFR estimates for Pennsylvania, from a single
  replication.}
\label{fig:retro-curves}
\end{figure}

Figure \ref{fig:retrospective-comparison}A disaggregates the MAEs by population
size, across the 51 regions. Observe that the spread of improvement varies by
population, with deconvolution generally having stronger gains over the
benchmarks for smaller population sizes (where estimation is generally more
challenging). Figure \ref{fig:retrospective-comparison}B confirms that this
holds for all orders of trend filtering regularization. 

Figure \ref{fig:retro-curves} shows the HFR curves for an example region,
Pennsylvania. The ratio-based estimates oscillate around the ground truth HFR
(solid black line), with the lagged ratio displaying clearly greater volatility, but
the convolutional ratio still showing a nontrivial amount as well.  Broadly, the
estimates from deconvolution do not share this behavior and are qualitatively
more stable. We emphasize that this is the case even though the ratio-based
methods use cross-validation with the 1se rule to tune their hyperparameters
(which incentivizes a greater amount of regularization). The same behavior is 
generally consistent across all geographies, and the HFR curves for the
remaining states are displayed in Figure 8 of Appendix
E of the Supplementary Material.

\subsection{Real-time analysis}

Deconvolution also performs strongly in the real-time setting. In the format
of Table \ref{tab:retro-cv} above, Table \ref{tab:rt-cv} reports the MAEs and 
percentage improvements over the benchmarks in the real-time case, when all 
hyperparameters are tuned with cross-validation. Again, for each method we
selected among the min and 1se rule depending on which resulted in more
favorable performance; in the real-time case, this ends up being the 1se rule
for all hyperparameters except for $\lambda$ in deconvolution, which resulted
in marginally better performance when tuned via the min rule. We note that the
convolutional ratio estimator had substantially worse performance under the min
rule, whose chosen window sizes tended to undersmooth counts. 

As we can see from the table, the deconvolution methods provide an improvement
in accuracy over the ratio-based methods, consistent with the previous results
for the retrospective case. All orders $m$ of trend filtering regularization are
around 11-17\% more accurate than the convolutional ratio, and around 24-29\%  
better than the lagged ratio. Tuning the order of trend filtering still
generally works well.        

\begin{figure}[tbp]
\centering
\includegraphics[width=0.82\textwidth]{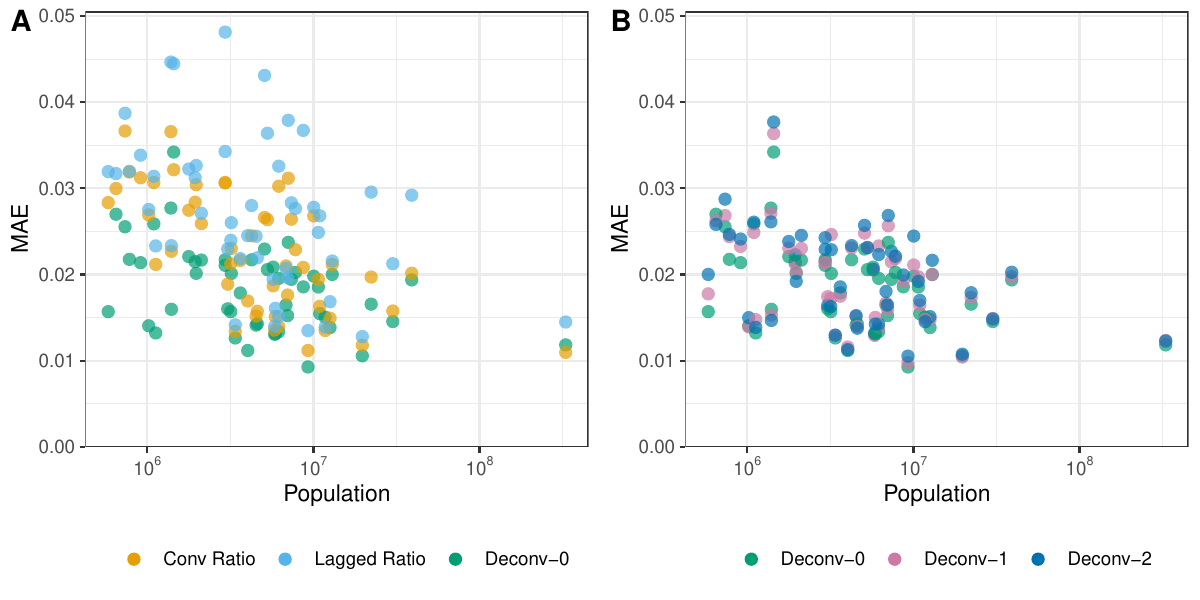}
\caption{MAE versus population size in real-time HFR estimation, for 51
  regions (\US\ and 50 states). Each point is averaged over 10 replications.}
\label{fig:rt-comparison}
\end{figure}

\begin{figure}[tbp]
\centering
\includegraphics[width=0.82\textwidth]{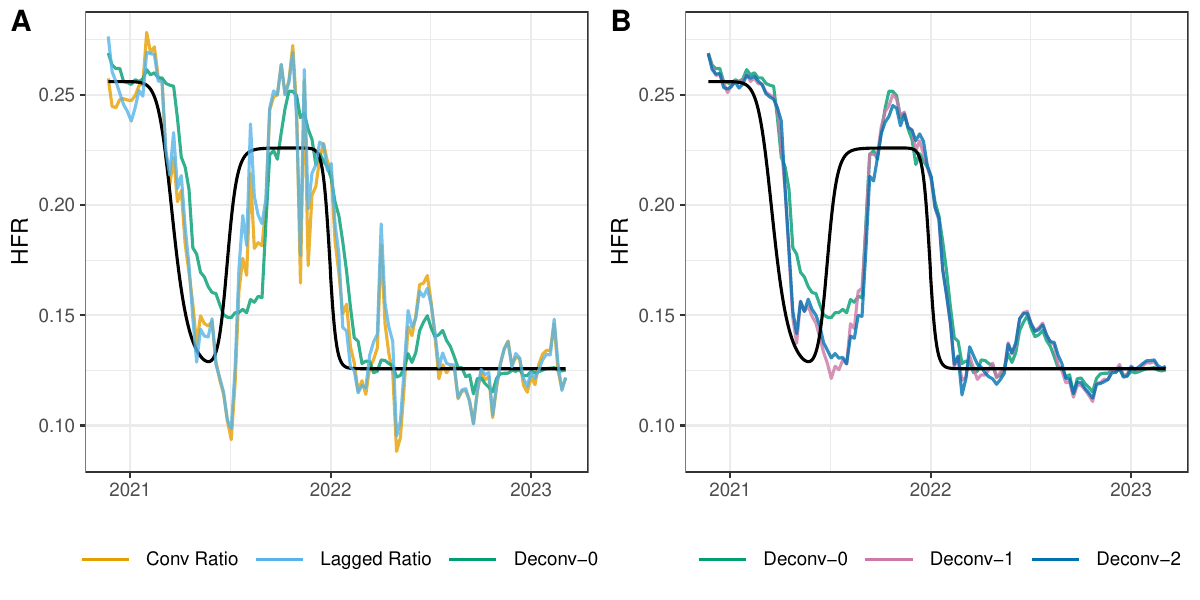}
\caption{Example real-time HFR estimates for Pennsylvania, from a single
  replication.}
\label{fig:rt-curves}
\end{figure}

Figure \ref{fig:rt-comparison} breaks down the MAE by population size, and once
again we can see that deconvolution has an advantage over the ratio methods
across the full range of geographies, with the gap being generally larger for
smaller regions. Figure \ref{fig:rt-curves} displays example HFR curves for
Pennsylvania, as before. In the current real-time case, we can clearly see the 
quality of all estimates degrade. The ratio-based estimates are especially
volatile, and their erratic behavior here can be understood from the
perspective of the analysis in \citet{goldwasser}. For example, their positive
bias in early 2022, particularly that of the lagged ratio, can be attributed to
the Omicron surge which just passed. The deconvolution estimates are
comparatively smoother and more stable. Finally, Figure 9
in Appendix E of the Supplementary Material displays the full set of HFR curves across all
states, where broadly the same conclusions are upheld. 

\begin{table}[tbp]
\centering
\caption{MAE of methods in real-time HFR estimation, and the associated
  percentage improvement on the convolutional ratio (CR) and lagged ratio
  (LR). Results are averaged over 51 regions and 10 replications.}     
\label{tab:rt-cv}
\begin{tabular}[t]{@{}lrrrrrr@{}}
\toprule
& Lagged Ratio & Conv Ratio & Deconv-0 & Deconv-1 & Deconv-2 & Deconv-Tuned\\
\midrule
MAE $\times \; 10^3$ & 27.0 ± 0.1 & 22.9 ± 0.1 & 18.5 ± 0.2 & 19.3 ± 0.1 & 19.8 ± 0.1 & 19.6 ± 0.1\\
Improv over CR (\%) & -19.3 ± 0.4 & 0.0 ± 0.0 & 16.7 ± 0.5 & 13.4 ± 0.4 & 10.9 ± 0.5 & 12.2 ± 0.5\\
Improv over LR (\%) & 0.0 ± 0.0 & 13.8 ± 0.3 & 28.6 ± 0.5 & 25.6 ± 0.5 & 23.6 ± 0.5 & 24.6 ± 0.5\\
\bottomrule
\end{tabular}
\end{table}

\subsection{Misspecification analysis}

\begin{figure}[t]
\centering
\includegraphics[width=0.95\textwidth]{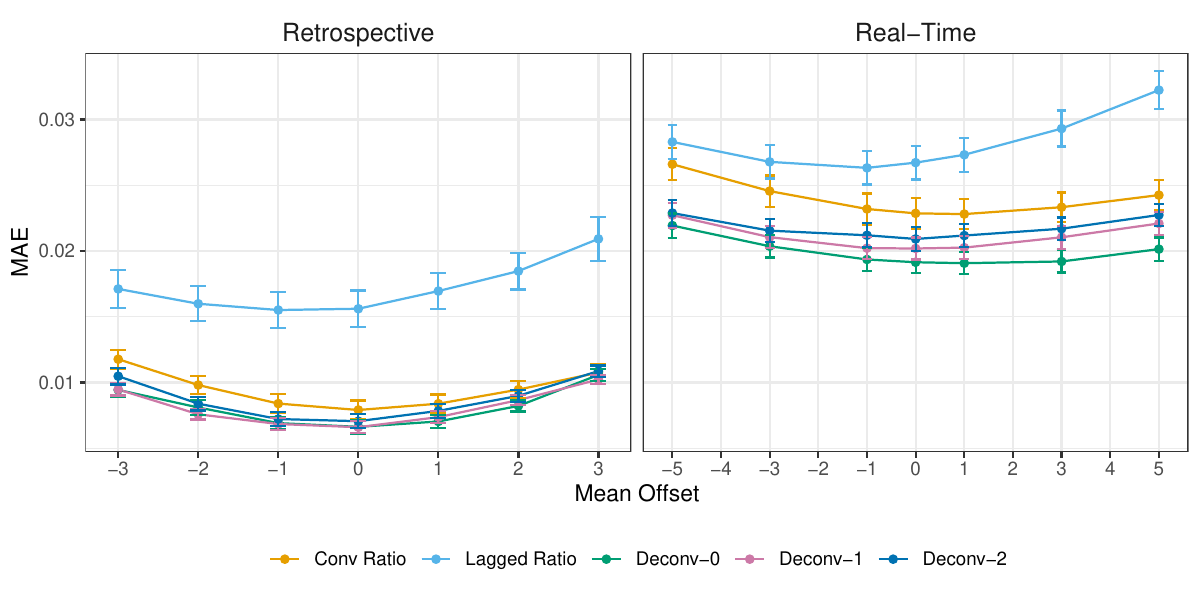}
\caption{MAE versus mean offset, which is a measure of misspecification of 
  the delay distribution. Results are averaged over 51 regions.}  
\label{fig:misp-maes}
\end{figure}

The advantage of deconvolution over benchmark methods persists under various
degrees of misspecification. Figure \ref{fig:misp-maes} displays the MAE as a
function of the offset of the mean of the working delay distribution used by
each method (or the lag used by the lagged ratio) relative to the true
delay. Unsurprisingly, we see accuracy degrade under misspecification, as
evidenced by the curves which generally slope upwards as the mean offset moves
away from zero. The gap in MAE of the lagged ratio and any of the other
estimators is clearly large, regardless of the amount of misspecification.
Comparing the convolutional ratio to deconvolution, the MAE gap narrows somewhat
as the mean offset grows, in the retrospective case; on the other hand, the gap
more or less holds steady as the mean offset varies, in the real-time case.
Figure 2 in Appendix E of the Supplementary Material transforms the MAE curves
from deconvolution in Figure \ref{fig:misp-maes} into percentage improvement
curves over the convolutional ratio. The findings are consistent and overall the
deconvolution methods display comparably strong performance in our misspecified
experiments.

  As a further robustness check, Appendix~E.5 of the Supplementary Material examines a
  time-varying delay distribution defined as a daily mixture of
  variant-specific gamma delays, with weights determined by each state's
  dominant SARS-CoV-2 variant on a given day. Deconvolution's advantage over
  the ratio-based benchmarks is broadly preserved in that setting, under both
  well-specified and misspecified working delays.

\section{Discussion}\label{sec:discussion}

In this work, we characterize the probabilistic relationship between two time
series (primary and secondary event counts) which are related by time-varying
convolution: a delay distribution times a quantity called the severity rate. The
resulting likelihood allows us to construct new deconvolution-based estimators
of severity rates, in both the retrospective and real-time settings. To
encourage smoothness over time, we apply trend filtering regularization, and in
the real-time case, we use additional regularization to control volatility at
the boundary of observations (the right tail of the time series). In extensive
simulations, designed to mimic the relationship between COVID-19
hospitalizations and deaths in the \US, we find our deconvolution method to be
consistently more accurate than the typical ratio-based methods used in
practice, often by a significant margin. On real data, the qualitative
differences are quite similar, with deconvolution able to provide stable
estimates of the hospitalization fatality rate (HFR) under changing conditions,
while the ratio-based estimates are subject to suspicious swings and spikes, as
expected based on \citet{goldwasser}.

It is worth further reflecting on some of our results from the perspective of
public health practice. In our real-time experiments, cross-validation and the
1se rule---to choose the smoothing window in the lagged ratio and convoluational 
ratio methods---nearly always selects the maximum window size of 4 weeks. Under
smaller window sizes, the ratio-based methods perform much worse and the margin
of improvement of deconvolution only grows. This creates a tradeoff to be
navigated, as long windows may be undesirable in practice. A core purpose of
real-time estimation is to expeditiously detect changes in severity
rates. Smoothing values over an entire month seriously hampers our ability to do
so, and it seems window sizes this large are rarely used in practice. The
ability of trend filtering to perform locally adaptive smoothing---which
effectively uses a longer or shorter smoothing window as needed at different
parts of the time series---is a potentially powerful way to navigate the basic
tradeoff between detecting changes and controlling variability.

We now reflect on some fundamental differences in the models underlying
ratio-based and deconvolution-based methods. Notice that, as defined in
\eqref{eq:severity}, severity rates are inherently forward-looking
quantities. The rate $p_t$ is defined not only by primary events at $t$, but
also by secondary events in the future. By maximizing an approximate likelihood,
the deconvolution-based method thus automatically adopts this forward-looking
perspective, even in the real-time case where future events are unobserved.  In
contrast, the real-time lagged and convolutional ratios from Section
\ref{sec:methods} are backward-looking methods. They quite literally looks
backwards in time to past primary events in order to explain secondary events at
$t$. To help draw this out precisely, we can define an analogous estimand, the
backward-looking severity rate, by  
\begin{equation}
\label{eq:back}
\tilde{p}_t = \sum_{k=0}^d \P(\text{secondary event occurs at $t$} \given 
\text{primary event at $t-k$}). 
\end{equation}
\label{eq:forw}
The original forward-looking severity rate admits a similar decomposition:  
\begin{equation}
p_t  = \sum_{k=0}^d \P(\text{secondary event occurs at $t+k$} \given
\text{primary event at $t$}).  
\end{equation}
In Appendix F of the Supplementary Material, we develop connections between the real-time
convolutional ratio and the backward-looking severity rate. A key point to
highlight: if conditions remain constant---meaning, the delay distribution and
severity rates are constant in a sufficiently large widow around $t$---then the
forward- and backward-looking severity rates are equal, \smash{$\tilde{p}_t =
  p_t$}, and the convolutional ratio is unbiased for this common value.    
Indeed, the real-time convolutional ratio performs well when the true severity
rates are nearly constant. Looking back to Figure \ref{fig:rt-curves}, its
estimates tend to be most accurate during such periods, for example, in the last
months of 2021, and throughout most of 2022. 

However, when severity rates are nonstationary, the convolutional ratio
suffers from bias. As shown in \citet{goldwasser}, this bias can be explained by
studying the differences between the backward-looking rate \smash{$\tilde{p}_t$}
and forward-looking rate $p_t$. In part, the success of the deconvolution-based
methods is due to the fact that they target the true foward-looking rate $p_t$,
although it is challenging to precisely characterize the resulting improvement
in bias for these methods. Of course, another significant component of their
success is the regularization employed, which improves variance. 

Interestingly, these forward- and backward-looking perspectives are mirrored in
the literature on reproduction numbers, a central quantity in computational
epidemiology. In broad terms, reproduction numbers convey the average number of
secondary infections that will be produced from a single primary infection.
\citet{fraser2007} uses the name ``case reproduction number'' for the forward
perspective, measuring the effect of an infection at time $t$ on future
infections. In contrast, the ``instantaneous reproduction number,'' often
denoted as $R_t$, is based on the backward perspective using secondary
infections occurring at $t$ due to primary infections that happened in the
past. As with severity rates, the case and instantaneous reproduction numbers
are equivalent under local stationarity. Most methods that estimate $R_t$ in
real-time target the instantaneous reproduction number \citep{Bettencourt2008,
cori2013new, Parag2021}.  Recent work by \citet{rtestim} proposes estimating
instantaneous $R_t$ in retrospect with trend filtering. This method could be
adapted to the real-time setting using our additional regularization
techniques. In addition, future work on $R_t$ could estimate case reproduction
numbers in real-time, using insights from our work on forward-looking severity
rates. Case reproduction numbers may be more desirable as estimands than their
instantaneous counterparts: they describe future dynamics of an epidemic in the
most direct and intuitive way.

We close by mentioning several avenues for future work. A key component in
practice is the delay distribution, whose misspecification can lead to
degradation in performance. Without high-quality line list data, obtaining an
accurate plug-in estimate is challenging. Jointly estimating this distribution
along with the severity rates themselves would evade this challenge, posing an
open methodological challenge. \citet{li2023reconstructing} examine a joint
estimation strategy for a different but broadly related deconvolution problem,
based on the expectation-maximization (EM) algorithm. It would be interesting to
study an analogous approach in our problem setting. 

Another interesting direction would be to estimate severity rates disaggregated
by different demographic groups, for example, severity rates for different age
brackets. Conceivably, this could be done with the same primary or secondary
event data (not requiring counts per demographic group) by employing a
cohort-based model. This would feature a common set of severity rate curves  
(per demographic group) across all regions, and would then mix by known
demographic proportions in each region to explain the effective severity rate in
that region. 

To our knowledge, no analogous Bayesian methods exist for the general class of severity rates we consider. However, a Bayesian formulation is a natural direction for future work, where prior specification could serve as an alternative regularization mechanism. 
Bayesian methods have been developed for related epidemiological quantities, including growth rates  \citep{guzmanrincon2023bayesian} and reproduction numbers \citep{cori2013new}.
In addition, several Bayesian works on compartmental models \citep{flaxman2020, gibson2023, korolev2020identification} place a prior on the IFR and recover a posterior distribution. 
It would be interesting to compare these IFR estimates to our frequentist framework's, after first deconvolving estimates of latent infection counts.


A final direction would be to produce confidence
intervals for severity rates obtained using deconvolution and trend
filtering penalties. In general, developing rigorous inferential tools on top of
trend filtering remains a challenge, but promising recent developments in data
fission/thinning \citep{leiner2025data, dharamshi2025generalized} suggest there  
may be a path forward. 

\section*{Significance Statement}

Public health authorities use severity rates to track the deadliness of a disease over the course of an epidemic. Typically, metrics like the case-fatality are estimated by dividing the number of new deaths by the number of recently reported infections. However, recent work reveals that this simple calculation can be seriously misleading, which poses large problems for decision-making during epidemics. When severity rates are underestimated, officials can miss genuine rises in risk and postpone critical interventions, and thus more people in harm’s way. Conversely, when these metrics are overestimated, authorities may enact needless restrictions and stoke undue public fear, wasting resources and eroding trust.

To address this, we propose methods to estimate time-varying severity rates, both in retrospect and real time. Our methods maximize a novel characterization of the likelihood, deconvolving the time series of severity rates that relate primary and secondary events (e.g. cases and deaths). The likelihood is regularized by a trend filtering penalty, which fits piecewise polynomials with adaptively selected knots in order to balance smoothness with strong local adaptivity. 

On a range of experiments, our methods consistently outperform the two leading benchmarks, reducing mean absolute error by roughly 15\% and 55\%. Moreover, they exhibit stronger qualitative behavior, with both smoother fits and less bias. These results support our class of estimators as a promising new approach to track public health risks as they unfold.

\begin{acks}[Acknowledgments]
The authors would like to thank the Associate Editor and anonymous referees
for their constructive comments that improved the quality of this paper.
\end{acks}

\begin{funding}
This work was supported by CDC and CSTE grant NU38FT00005.
\end{funding}

\begin{supplement}
\stitle{Supplement to ``Estimating Time-Varying Epidemic Severity Rates with
  Adaptive Deconvolution''}
\sdescription{The supplementary document contains the proof of
  Proposition~\ref{prop:corr}, additional derivations, and further
  experimental details and results referenced throughout the paper.}
\end{supplement}

\begin{supplement}
\stitle{Code and data}
\sdescription{A zip archive containing all code and data files used to
  generate the analyses, figures, and tables in the paper.}
\end{supplement}

\bibliographystyle{imsart-nameyear}
\bibliography{refs}

\end{document}